# Model Based Sensor System for Temperature Measurement in R744 Air Conditioning Systems

Sven Reitz[1], Andreas Schroth[2], Peter Schneider[1]
[1]Fraunhofer Institute for Integrated Circuits, Division Design Automation, Dresden, Germany
[2]Intelligente Sensorsysteme Dresden GmbH, Dresden, Germany
Email: sven.reitz@eas.iis.fraunhofer.de, Tel: +49 (351) 4640 840

*Abstract* - The goal is the development of a novel principle for the temperature acquisition of refrigerants in $CO_2$ air conditioning systems. The new approach is based on measuring the temperature inside a pressure sensor, which is also needed in the system. On the basis of simulative investigations of different mounting conditions functional relations between measured and medium temperature will be derived.

## I. INTRODUCTION

The number of sensor systems for monitoring and control tasks in the automotive industry, automation technology and medical technology has been highly increased by the ongoing technological development in the past years. Powerful sensors for the precise detection of process parameters such as pressure, temperature and flow play a key role in the realization of innovative system solutions.

The goal is the development of a novel principle for the temperature acquisition of refrigerants in $CO_2$ air conditioning systems. This type of air conditioners will be used in the automotive sector medium to long term, based on the current ongoing discussions on new EU regulations about banning on previously used refrigerants based on hydro fluorocarbons [1, 2].

Essential requirements for the temperature measurement are a fast response behavior of the sensor and a preferable small impact to the flow of refrigerant. Both requirements are contradictory, since to reach a short response time a direct positioning of the temperature sensor in the refrigerant flow is useful. This is necessarily associated with a reduction of the cross-section, disturbing the flow of refrigerant. Another problem is that the sensor housing dissipates heat and this causes a distortion of the measured values.

The new approach is based on measuring the temperature inside a pressure sensor, which is also needed in the system [3]. The temperature sensor will be outside the refrigerant flow and therefore the measured temperature will differ from the actual medium temperature. On the basis of simulative investigations of different mounting conditions correction maps will be determined. These maps, stored in the intelligent evaluation electronics of the system, are used for computation of the actual temperature of the medium.

Fig. 1 shows the typical mounting condition of the pressure-temperature sensor.

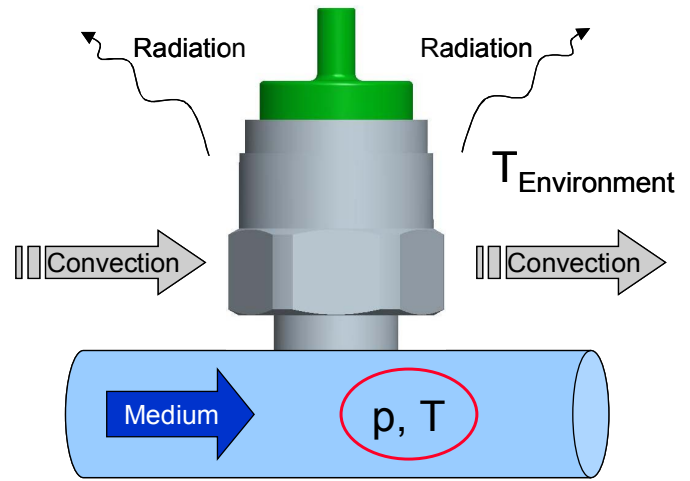

Fig. 1. Schematic representation of the pressure-temperature sensor

## II. METHODOLOGY

The challenge is that the temperature of the medium directly in the tube has to be measured but the appropriate sensor can only be implemented in a certain distance from the desired measuring point (Fig. 2). Furthermore, it is characteristic that the temperature at the sensor point may deviate significantly from the temperature of the medium especially due to thermal convection and radiation (Fig. 1).

An approach was developed, which is based on an approximation of the medium temperature at position x from measurements at one or more locations $y_1$, $y_2$, $y_3$, ... (Fig. 2). In order to determine valid relations and characteristic diagrams

$$T(x) = f(T(y_1), T(y_2), T(y_3), \ldots) \quad (1)$$

for a specific application and design, simulation runs can replace complex or technically not feasible measurements [4, 5]. Based on the characteristic diagrams determined by simulations, for example multidimensional regression can be used to derive the above mentioned functional relation (1). The correction calculation itself will be done in the sensor system (for example by means of integrated microcontrollers), or externally in a higher-level control unit.

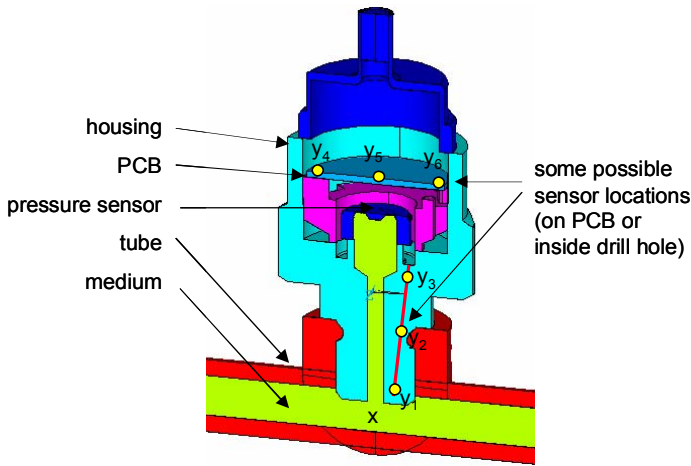

Fig. 2. Half-model of the sensor due to symmetry in ANSYS

For the model-based acquisition of the relevant physical effects and the essential boundary conditions of the sensor system it was necessary to create ANSYS [6] finite-element models (FE models) of the existing pressure-temperature sensor including its connection to the medium and the environment. The models were created in such a manner that they include the heat conduction and capacity of the used materials as well as take the heat convection and radiation to the environment (air) into account. Furthermore, the symmetry of the assembly was exploited to reduce the problem to a half-model (Fig. 2).

The detailed FE-models will be used to investigate
- geometry variations of the sensor system
- different positions of the temperature sensors and
- the influence of mounting and ambient conditions,

and thus they serve for the precise determination of the required characteristic diagrams.

## III. COMPREHENSIVE SIMULATION STUDIES

As a first step static simulations were carried out. Medium and ambient temperature, convection and radiation were assumed to be constant. On the one hand, the resulting temperature distribution in the entire sensor system can be determined (Fig. 3 left), on the other hand, the temperatures at the selected 6 measurement points are a direct result of the simulation (Fig. 2). In the next step, by varying the medium and ambient temperatures characteristic diagrams, as shown in Fig. 3 right for the 3 sensors in the drill hole exemplary, were obtained. With the help of these maps the relationship between the temperatures at the sensor locations and the medium temperature had to be found.

## IV. RELATION BETWEEN SENSOR TEMPERATURES AND MEDIUM TEMPERATURE

The optimal number and location of the at most 6 temperature sensors (Fig. 2), necessary for an approximation of the medium temperature, had to be found.

First, for different combinations of temperature sensors approximation functions for the medium temperature based on characteristic diagrams were derived. Afterwards, the resulting differences between calculated and given medium temperature were investigated.

Essential for a most accurate approximation is the choice of suitable approach functions. From the multitude of approach functions and the use of 2 to 6 temperature sensors, the following approach

$$T = b_0 t_1 + b_1 t_1 t_1 + b_2 t_5 + b_3 t_5 t_5 + b_4 t_1 t_5 \qquad (2)$$

should be mentioned. The medium temperature is calculated by this function using the lowest temperature sensor in the drill hole ($t_1$) and the middle sensor on the printed circuit board ($t_5$) (Fig. 2). These two sensors were selected because the lowest sensor in the hole is very close to the medium and the sensor on the printed circuit board is relatively close to the environment. Using this approach, even in practical temperature measurements the resulting differences between given and calculated medium temperatures are below 1.8 K.

## V. IMPLEMENTATION OF APPROXIMATION FUNCTION

For the implementation of the approximation functions a microcontroller board equipped with an 8-bit microcontroller from Atmel (ATmega128) was used as an experimental platform (Fig. 4). This controller is connected over two analog inputs to the temperature sensors and a serial

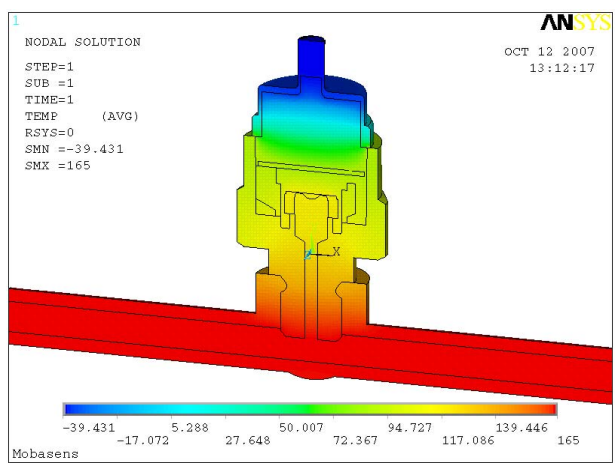
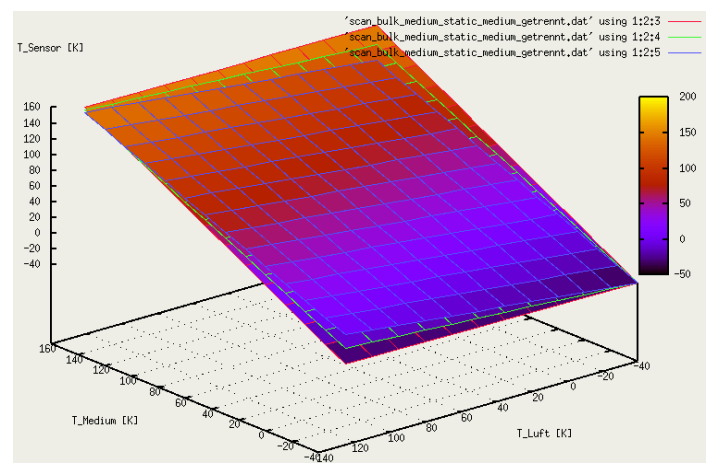

Fig. 3. left: Temperature distribution of entire sensor system,
right: Temperatures at sensor locations in drill hole as function of medium and environment temperatures

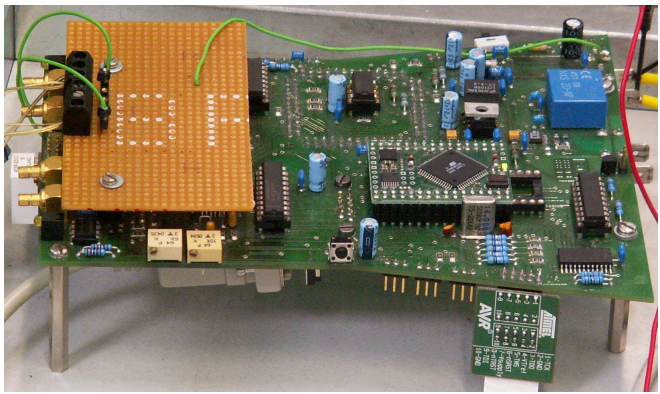

Fig. 4. Test platform using an AVR microcontroller (ATmega128) for data acquisition and medium temperature calculation

interface is used to transmit the calculated medium temperature to a PC.

A block library for Matlab/Simulink, which was developed at the Institute, allows the configuration of the required peripherals on the test board and generates C-code automatically, e.g. for ATmega128. At first, the Simulink model was verified by simulation runs, and then automatically transferred to the board. Using this hardware solution the medium temperature was calculated successfully.

The program code generated from the Simulink model will also be used for an estimation of the required memory resources of a microcontroller assembled in future sensor systems. For the setup described above, the controller uses 10 kB of Flash, 320 Bytes of RAM and runs with sample rates up to 1 kHz.

## VI. CONCLUSIONS

The given results show that simulative studies of the thermal behavior of the sensor system lead to a correlation between the temperatures at the possible sensor positions and the medium temperature. Furthermore, variations in the design of the sensor system and the thermal boundary conditions are much easier to investigate than in the real system.

Comparisons between simulation and measurement results have shown a very good accordance of the finite element model with the real sensor system.

The use of two temperature sensors in the sensor system is sufficient to compute the medium temperature in the case of known thermal convection (i.e. known mounting conditions in practice). The effect of changing thermal convection and radiation respectively is subject of further investigations.

The derived approach and approximation functions can be described in behavioral description languages (e.g. Modelica [7, 8]) or Matlab/Simulink.

Using system-level tools like Matlab/Simulink for design, simulation (Verification) and code generation has proven to be very time efficient and still allows the algorithm to be implemented in very small and cost effective microcontrollers.

## ACKNOWLEDGMENT

This paper is based on the project MOBASENS which is supported by the Sächsische Aufbaubank - Förderbank - under support-no. 11939/1925. The authors of this paper are solely responsible for its content.